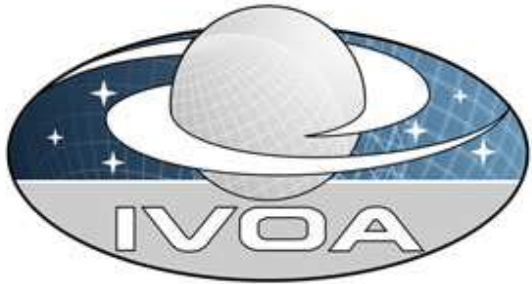

*International*

*Virtual*

*Observatory*

*Alliance*

# Maintenance of the list of UCD words

# Version 1.2
## *IVOA Recommendation  28 May 2006*

**This version:**
  http://www.ivoa.net/Documents/REC/UCD/UCDlistMaintenance-20060528.html
**Latest version:**
  http://www.ivoa.net/Documents/latest/UCDlistMaintenance.html
**Previous version(s):**
  http://www.ivoa.net/Documents/PR/UCDmaintenance/UCDlistMaintenance-20060421.html
  http://www.ivoa.net/Documents/PR/UCDmaintenance/UCDlistMaintenance-20060307.html
  http://www.ivoa.net/Documents/WD/UCDmaintenance/UCDlistMaintenance-20051107.html

**Editor(s):**
   A. Preite Martinez, S. Derriere
**Author(s):**
   Nausicaa Delmotte,
   Sebastien Derriere,
   Norman Gray,
   Robert Mann,
   Jonathan McDowell,
   Thomas Mc Glynn,
   François Ochsenbein,
   Pedro Osuna,
   Andrea Preite Martinez,
   Guy Rixon,
   Roy Williams

## Abstract

According to what is stated in the IVOA Recommendation *An IVOA standard for Unified Content Descriptors*, a procedural document should be created in order to maintain (add, change, suppress words) the standard list of UCD1+ words *The UCD1+ Controlled Vocabulary* . This document describes the procedure to maintain the standard list of UCD1+ words.

## Status of This Document

This is an IVOA Recommendation. The first release of this document was 2006 May 28.

*This document has been produced by the IVOA UCD Working Group.*

*A list of current IVOA Recommendations and other technical documents can be found at http://www.ivoa.net/Documents/.*

## Acknowledgements

We acknowledge the support of the VO-Tech project.

## Contents

# 1   Introduction

According to the IVOA Recommendation *An IVOA standard for Unified Content Descriptors*, the present document describes in detail the procedure to add, modify or suppress UCD-words in the standard list *The UCD1+ Controlled Vocabulary*. The modification of the standard list of UCD words is conceived as a flexible but controlled process. A Scientific Board, as defined by the UCD (IVOA-Rec) document above, has the role of studying the proposed modifications and, within a given period of time, to give an answer as to whether such modifications can or cannot be accepted.

This document addresses:
- the contact point to ask for a modification of the UCDs
- the life-cycle of the process of modification of UCDs
- when and how a new UCD becomes part of the standard.

These actions are supported by an automatic web-based form that has to be filled in and sent to the UCD Scientific Board (**ucd-sci@ivoa.net**), giving an answer back to the user acknowledging the request, and giving a time estimate for an answer.

The parties involved in the procedure are:
- A **user** submitting a modification: e.g. any member of the IVOA community.
- The **Scientific Board**: open to the IVOA community. Formed by, and under the responsibility of, the chairman of the UCD/Semantics WG. The Board is coordinated by a moderator, proposed by the chairman of the WG. The responsibility of the Board consists in studying the proposals of modifications of UCD words, and to figure out whether the proposal should be accepted or rejected.
- The **UCD/Semantics WG**: The role of the chairman of the WG is that of supervising the work of the Scientific Board, and to collect at given times the approved modification into a draft new version of the UCD list. The members of the WG will discuss the new version of the UCD list.
- The **IVOA Community**: The document approved by the UCD/Semantics WG will then follow the standard IVOA procedure to reach the Recommendation status.

# 2   Procedure to Modify UCD Words

The procedure complies with the following basic principles:

- Any member of the IVOA community can submit at any time a request for modification (RFM) of the standard list of UCD words;
- An automatic web-based form is the interface between the community and the Scientific Board in charge of discussing the RFM;

- A deadline for concluding the discussion and for answering the RFM should be given to the author of the RFM;
- At regular intervals (synchronized with InterOp meetings) the approved modifications will be integrated in a new version of the standard UCD1+ Common Vocabulary.

The procedure is described in detail in the following sub-sections.

## 2.1   Request for Modification (RFM)

The first step in the process of modification of the standard list of UCD-words is a request for modification (RFM) that any member of the IVOA community can present at any time. The requested modification of one or more UCD-words should comply with the rules defined in the document "An IVOA standard for Unified Content Descriptors" for building UCD-words, and should be submitted as indicated in section 2.2.

An RFM can consist of a request for:

- *Deletion*: The UCD word is to be removed from the list of standard UCDs. Approval of a deletion request should include a statement of what UCD would be used in place of the original UCD. A list of deleted UCDs and their replacements should be maintained.
- *Amendment*: The description of the UCD and/or its syntactic role (primary, secondary, etc.) and/or its spelling is to be corrected. Amendment of the spelling of a UCD word corresponds to a *deletion* (of the word with the old spelling) followed by an *addition* (see below) of the word with the new spelling. A request to better clarify the description of an UCD word that is considered too short or cryptic or ambiguous, and/or to change the syntactic role of the UCD word, is called *clarification*.
- *Addition*: A new UCD word is to be added.  In addition to the general description, there needs to be an indication of the syntactic role, which determines  how this word can be combined with other words (see the list of syntax codes in Appendix A of the document *The UCD1+ Controlled Vocabulary* ).

## 2.2   Web-based form for submitting RFMs

A web-based form for requesting a modification of UCD words has been set up and it is accessible from the IVOA WG page at:
http://www.ivoa.net/twiki/bin/view/IVOA/IvoaUCD .
The form is composed of at least 6 fields:

- UCD: The UCD-word involved

- Definition: The description of the word in common language
- Remarks: Comment or rationale of the RFM
- Suggested by: The name of the submitter
- Date: The date of submission
- Answer: The answer to the RFM

The first 4 fields will be filled up by the submitter. The date is inserted automatically by the form. The last field will contain the answer to the RFM written by the moderator of the Scientific Board at the end of the discussion within the board. It is the responsibility of the moderator to respond to all RFMs, summarizing the discussion in the board.
The form will be sent by e-mail to all the members of the Scientific Board as soon as the RFM is submitted.
The notification will contain all the fields concerning the new submission.

## 2.3  Evaluation of RFMs

All RFMs will be open to discussion and will be evaluated by the Scientific Board. A two-week period for discussion is automatically allowed, starting from the day of notification to the members of the board. Extensions of this period can be decided by the moderator of the board, with notification to the user/submitter of the RFM and to the Scientific Board.
At the end of the discussion, the moderator will answer the RFM using the $6^{th}$ field of the form. The form will send an e-mail to the user, the Scientific Board and the chairman of the [UCD/Semantics WG](#).
A negative answer must be explained.
The moderator of the Scientific Board can suggest the implementation of a period of collection of RFMs in order to group together a more significant number of them, if this improves the efficiency of the discussion within the Board.

## 2.4  Maintenance of the UCD Controlled Vocabulary

At least every year, but usually synchronized with InterOp meetings, the chairman of the UCD/Semantics WG will edit the last approved version of the document "The UCD1+ Controlled Vocabulary" in order to include the newly approved RFMs, update the version, and change the status of the document to IVOA Working Draft.  The draft will be open to discussion within the WG for a length of time consistent with the importance/number of modifications, but no longer than 4 weeks.
At the end of the discussion within the WG, the document will follow the IVOA procedure (PR, RFC, Exec vote, Rec) for document standards [1] [2].
An historical record of all approved modifications will be included in the document.
If necessary, the chairman of the WG can decide a different cadence of maintenance of the UCD list, not necessarily synchronized with InterOp meetings.

In case of explicit request by the user for immediate use of an approved new UCD-word, the chairman of the WG, after the approval of the RFM by the Scientific Board and at the end of the discussion in the WG, can authorize the use of that word with the temporary namespace "**tmp:**". In such a case the IVOA community (interop@ivoa.net) shall be notified by the chairman of the WG.

In the case of RFMs that only deal with special amendments of the form of *clarifications* (see section 2.1) of the description of UCD words, the process of maintaining the document "The UCD1+ Controlled Vocabulary" is simplified. After the discussion of the requests for clarification in the Scientific Board and in the UCD/Semantics WG, the chairman of the WG will include the accepted clarified descriptions in a new version of the document, with an historical record of the approved modifications.

## References

[1] R. Hanisch, *Resource Metadata for the Virtual Observatory* , http://www.ivoa.net/Documents/latest/RM.html
[2] R. Hanisch, M. Dolensky, M. Leoni, *Document Standards Management: Guidelines and Procedure* , http://www.ivoa.net/Documents/latest/DocStdProc.html

## A. Changes from previous versions

Changes from version 1.1 (PR 2006 March 07):

1. List of authors upgraded

2. Table of content: added "Changes from previous versions"

3. pag.3-4, par.1-2: changed "community" into "IVOA community"

4. pag.3, par.2: changed "dead-line" into "deadline"

5. pag.4, par.2.1: changed "cryptical" into "cryptic"

6. pag.4, par.2.1: change in paragraph *Amendment*:
    sentence with definition of *clarification* moved at the end of paragraph

7. pag.4,par.2.1: changed paragraph *Addition* as follows:
    "*Addition*: A new UCD word is to be added. In addition to the general description, there needs to be an indication of the syntactic role, which determines how this word can be combined with other words (see the list of syntax codes in Appendix A of the document The UCD1+ Controlled Vocabulary)."

8. pag.4, par. 2.2 beginning changed as follows:

"A web-based form for requesting a modification of UCD words has been set up and it is accessible from the IVOA WG page at:
http://www.ivoa.net/twiki/bin/view/IVOA/IvoaUCD .
The form is composed of at least 6 fields:"

9. pag.5,par.2.3: changed "A negative answer shall be motivated"
into "A negative answer must be explained"

9. pag.6: introduction of Appendix A – Changes from previous versions.